\documentclass[12pt,preprint]{aastex}
\usepackage{graphicx}
\usepackage{epsfig}
\usepackage{amsmath,amssymb}
\usepackage{lscape}
\usepackage{verbatim}
\usepackage{rotating}
\usepackage{subfigure}

\def\lapprox{\lower.4ex\hbox{$\;\buildrel <\over{\scriptstyle\sim}\;$}}
\def\gapprox{\lower.4ex\hbox{$\;\buildrel >\over{\scriptstyle\sim}\;$}}

\shorttitle{Missing Baryons}
\shortauthors{Anderson et al.}

\begin{document}
\title{Do Hot Haloes Around Galaxies Contain the Missing Baryons?}
\author{Michael E. Anderson\altaffilmark{1}, Joel N. Bregman\altaffilmark{1}\\}
\altaffiltext{1}{Department of Astronomy, University of Michigan, Ann Arbor, MI 48109; 
michevan@umich.edu, jbregman@umich.edu}

\begin{abstract}
Galaxies are missing most of their baryons, and many models predict these baryons lie in a hot halo around galaxies. We establish observationally motivated constraints on the mass and radii of these haloes using a variety of independent arguments. First, the observed dispersion measure of pulsars in the Large Magellanic Cloud allows us to constrain the hot halo around the Milky Way: if it obeys the standard NFW profile, it must contain less than 4-5\% of the missing baryons from the Galaxy. This is similar to other upper limits on the Galactic hot halo, such as the soft X-ray background and the pressure around high velocity clouds. Second, we note that the X-ray surface brightness of hot haloes with NFW profiles around large isolated galaxies is high enough that such emission should be observed, unless their haloes contain less than 10-25\% of their missing baryons. Third, we place constraints on the column density of hot haloes using nondetections of OVII absorption along AGN sightlines: in general they must contain less than 70\% of the missing baryons or extend to no more than 40 kpc. Flattening the density profile of galactic hot haloes weakens the surface brightness constraint so that a typical L$_*$ galaxy may hold half its missing baryons in its halo, but the OVII constraint remains unchanged, and around the Milky Way a flattened profile may only hold $6-13\%$ of the missing baryons from the Galaxy ($2-4 \times 10^{10} M_{\odot}$). We also show that AGN and supernovae at low to moderate redshift - the theoretical sources of winds responsible for driving out the missing baryons - do not produce the expected correlations with the baryonic Tully-Fisher relationship and so are insufficient to explain the missing baryons from galaxies. We conclude that most of missing baryons from galaxies do not lie in hot haloes around the galaxies, and that the missing baryons never fell into the potential wells of protogalaxies in the first place. They may have been expelled from the galaxies as part of the process of galaxy formation.

\end{abstract}


\maketitle

\section{Introduction}

The so-called ``missing baryon problem'' actually refers to two separate, but related, issues. First is the realization that the mean density of all detected baryons in the local universe accounts for less than half of the cosmological baryon density. The universal baryon fraction is known precisely from, among other sources, the Wilkinson Microwave Anisotropy Probe 5-year data (Dunkley et al. 2009): $f _b \equiv \Omega_b / \Omega_m = 0.171 \pm 0.006$ and $\Omega_b h^2 = 0.0227 \pm 0.0006$. But studies that attempt to count baryons in the low-redshift universe (Fukugita, Hogan, and Peebles 1998, Fukugita and Peebles 2004, Nicastro et al. 2005a) find that only about a tenth of this figure is observed in stars and cool gas in galaxies, while another third of $\Omega_b$ is divided between the hot plasma in the intracluster medium of galaxy clusters, and the cool intercluster gas detected in Ly$\alpha$ absorption lines (for more on the latter, see e.g. Penton, Stocke, and Shull 2004).

This result introduces the second missing baryon problem, which is the realization that most galaxies are severely baryon-depleted relative to the cosmological fraction (e.g. Bell et al. 2003). Based on cosmological simulations, the rest of the baryon budget is thought to reside in the  ``warm-hot intergalactic medium'' (WHIM), at temperatures between $10^5$ and $10^7$ K (Cen and Ostriker 1999, Dav\'{e} et al. 2001). This is thought to include most of the baryons associated with galaxies, since the optically-luminous portion of galaxies is severely baryon-depleted relative to the universal fraction (Dai et al. 2010). Observational evidence for the existence of the WHIM has subsequently begun to accrue, primarily through detections with ultraviolet and X-ray telescopes of highly-ionized Oxygen absorption lines (for a recent review, see Bregman 2007). 

These observations do not yet constrain the nature and distribution of the WHIM. Essentially all models agree that the WHIM exists as a cosmic filamentary web with some material in hot haloes around galaxies, but the mass fraction as a function of WHIM density is still unknown. Some numerical models (e.g Dav\'{e} et al. 2009) find a diffuse WHIM at $z=0$ with most of its baryonic content lying outside of galactic haloes. However, many simulations either assume (e.g. Bower et al. 2006, Croton et al. 2006) or self-consistently derive (e.g. Cen and Ostriker 2006, Tang et al. 2009, Kim, Wise, and Abel 2009) the result that cosmologically significant reservoirs of baryons are embedded in hot haloes around massive galaxies ($kT \sim $ a few hundred eV). These hot haloes have also been modeled in separate theoretical calculations (Fukugita and Peebles 2006, Dekel and Birnboim 2006, Kaufmann et al. 2009) and have important consequences in the galaxy formation process such as differentiating the ``blue cloud'' from the ``red sequence'' (Dekel and Birnboim 2006, Bouch\'{e} et al. 2009). Observationally, the evidence for hot haloes is still unclear: gas at these temperatures has been observed in X-rays around disk galaxies out to a few kpc beyond the disk (Strickland et al. 2004, Li et al. 2006, Li, Wang, and Hameed 2007), but this gas seems to result from starburst activity in the host galaxy instead of emerging as a byproduct of galaxy formation (Rasmussen 2009), and its inferred mass falls significantly short of theoretical predictions (T\"{u}llmann et al. 2006). 

In this paper, we examine the implications of this paradigm in more detail.  If most of the baryons associated with galaxies reside in hot haloes around the optically luminous part, there are several predictions that can be tested with existing observations. We seek to answer two questions. First, how much baryonic mass resides in these haloes? And, what can these haloes and the second missing baryon problem tell us about the process of galaxy formation? 

To answer these questions, we present several independent lines of argument. First, we consider observational constraints on the density of hot gas around the Milky Way using the dispersion measure of pulsars in the Large Magellanic Cloud, as well as the ambient pressure around high velocity clouds, and the galactic soft X-ray background. We then extend this argument to other galaxies. We constrain the characteristic density and radius of hot haloes using existing X-ray observations of absorption along quasar sightlines and emission from galactic haloes.  We also discuss the energetics of driving a galactic-scale wind, arguing that existing mechanisms for expelling most of the missing baryons seem incomplete. Finally, we discuss flattened density profiles and other prospects for resolving the second missing baryon problem. Before engaging these arguments, however, we will begin by describing the assumptions we use to model hot haloes, which are very similar to those of Fukugita and Peebles (2006).

\section{On Milky Way Parameters}
Much of the following discussion directly relates to the Milky Way or uses scaling relations to extrapolate from the Galaxy to other galaxies. Therefore we begin by discussing the parameters and scaling relations used throughout this paper.

Recent estimates for the virial mass of the Milky Way vary significantly. The consensus figure is roughly $2.5 \times 10^{12} M_{\odot}$ (Sakamoto, Chiba, and Beers 2003, Bellazzini 2004, Loeb et al. 2005, Li and White 2008) , although a recent estimate was as low as $1.0^{+ 0.3}_{-0.2} \times 10^{12} M_{\odot}$ (Xue et al. 2008) while another (Reid et al. 2009) revised upwards the circular velocity by about 15 km s$^{-1}$, which should increase the mass estimate by more than 10\%. Since larger masses for the Galaxy increase the amount of missing baryons, we will adopt a conservative virial mass estimate of $2.0 \times 10^{12} M_{\odot}$. 

Regarding the virial radius of the Milky Way, there is scatter around a central value. Shattow and Loeb (2009) use $R_{\text{vir}} = 277$ for a solar distance from the Milky Way center of $d_{\odot} = 8.0$ kpc, while the earlier work of Loeb et al. (2005) uses a virial radius of 207 kpc. Klypin, Zhao, and Somerville (2002) favor a virial radius of 258 kpc. We adopt an estimate for the virial radius of 250 kpc.

Another relevant Galactic parameter is amount of missing baryons. There is a surprisingly tight correlation between the total mass of a galaxy and its baryon fraction (McGaugh 2000, Dai et al. 2010), which we parametrize as $f_b = 0.04(M/2\times10^{12} M_{\odot})^{1/2}$ for $M \lapprox 6\times10^{12} M_{\odot}$. The relation flattens significantly for galaxy groups and clusters, but those are outside the scope of this work. This implies a baryonic mass of $8 \times 10^{10} M_{\odot}$ for the Galaxy, which broadly comports with the estimate of a stellar mass of $5 \times 10^{10} M_{\odot}$ for the Galaxy (Binney and Tremaine 2008), with an additional $1-2 \times 10^{10} M_{\odot}$ in gas and dust. We assume the total mass is comprised of three components: the dark matter halo, the observable baryons in the galaxy, and the missing baryons in a hot halo. In this model, the mass of the missing baryons can be related to the observed mass $M_0$ of the other two components:

\begin{equation}M_{\text{miss}} = \frac{0.17 - f_b}{0.83+f_b} M_0\end{equation}

For the Milky Way, the missing baryonic mass is therefore $3 \times 10^{11} M_{\odot}$. As a simple first-order model, we assume some fraction of the missing mass is distributed in a hot virialized halo around the disk. We can model the gaseous halo by assuming the Galactic dark matter halo obeys a Navarro, Frenk, and White (NFW) profile (Navarro, Frenk, and White 1997)). The missing baryons would follow the gravitational potential set by the dark matter, and therefore also obey the same density profile, except that the missing baryons do not extend into the disk, so we occasionally truncate the gaseous halo at an inner radius $r_1$ (this does not have a significant effect on our analysis). We examine many values for $r_1$ and for the NFW concentration parameter $C \equiv r_0 / R_{\text{vir}}$, but our nominal choice will be $r_1 = 10$ kpc and $C = 12$ (so $r_0 = 20.8$ kpc). 

Finally, the gaseous halo is assumed to be approximately isothermal at the virial temperature $k T_{\text{vir}} \sim 300$ eV. This is also the temperature of the hot gas observed around NGC 891 (T\"{u}llmann et al. 2006), a near 'twin' of the Milky Way (van der Kruit 1984), and is similar to the 250 eV gas surrounding the Galactic disk observed by Snowden et al. (1998). We expect the gas to have $Z \sim 0.2 Z_{\odot}$ (using Anders and Grevesse 1989 abundances)  following Cen and Ostriker (2006) and Rasumssen et al. (2009), and since similar metallicity is found in the hot intracluster medium even towards the virial radius of distant galaxy clusters (Maughan et al. 2008, Anderson et al. 2009). In computing the electron density of the halo, we take $\mu = 0.6$ and $\mu_e = 1.2$.

\section{Constraints on the Milky Way Halo}

\subsection{LMC Pulsar Dispersion Measure}

The strongest constraint we can derive for the hot halo around the Galaxy comes from the dispersion measure (DM) of pulsars in the Large Magellanic Cloud (LMC). The DM is defined as \\

\begin{equation} DM = \int_0^d n(l) \text{d}l\end{equation}

where $d$ is the distance to the pulsar and $n(l)$ is the free electron density along the line of sight. The Galactic free electrons (primarily from the so-called ``Warm Ionized Medium'', or ``WIM'') have been modeled as an exponential disk by, e.g, Gaensler et al. (2008). Integrating over this disk out to the LMC (and assuming a distance of 50 kpc from the Sun), we estimate the Galactic WIM contributes about 47 cm$^{-3}$ pc to the DM of pulsars in the LMC. In a survey of the Magellanic Clouds, Manchester et al. (2006) measured the DM of 11 pulsars in the direction of the LMC. Three pulsars have DMs below 47 cm$^{-3}$ pc (the values are 26,28, and 45 cm$^{-3}$ pc), suggesting these sources are located within the Galaxy and randomly superimposed in front of the LMC. The next-lowest DMs for LMC pulsars are 66, 69, and 73 cm$^{-3}$ pc, which leads to a conservative estimate of $\approx 70$ cm$^{-3}$ pc for the minimum DM introduced by the free electrons towards the LMC. This is also consistent with the lowest DMs for pulsars in the same survey towards the Small Magellanic Cloud (SMC): 70 and 76 cm$^{-3}$ pc. 

Subtracting 47 cm$^{-3}$ pc of Galactic electrons leaves 23 cm$^{-3}$ pc of electrons from other sources. The LMC has an ionized halo that may be responsible for much of this excess DM. One can assess the amount of ionized gas from the high-ionization absorption lines, such as OVI and CIV. The column densities of these ions in the LMC are comparable to the Milky Way (Howk et al. 2002). Considering that the metallicity is lower in the LMC, the inferred electron column is likely to be greater than that of the Milky Way. This implies that only a fraction of the excess DM is due to an ambient medium along the line of sight toward the LMC. However, since we do not know precisely the position of the LMC pulsars within its halo, in order to keep our estimate as conservative as possible we neglect the LMC electrons and assign all 23 cm$^{-3}$ pc of excess DM to the hot halo of the Milky Way. 

With an LMC distance of 50 kpc, 23 cm$^{-3}$ pc corresponds to an average electron density between the Sun and the LMC of $<n> \approx 5\times10^{-4}$ cm$^{-3}$.  If the entire halo had this mean density out to the virial radius, it would contain $8\times10^{11} M_{\odot}$ -- all of the missing baryons. This is an unphysical profile, however. We instead consider the standard assumption of an NFW profile for the missing baryons, with a concentration of 12 (Klypin, Zhao, and Somerville 2002), and find a halo obeying this profile and the DM constraint would have a mass of $1.2\times10^{10} M_{\odot}$, which is only 4\% of the missing baryons from the Galaxy. However, since the Galactic electrons have a scale length of $1.8$ kpc (Gaensler et al. 2008), we might reasonably argue that baryons within a few scale lengths are Galactic, not missing baryons. We therefore also consider an NFW profile with an inner truncation radius of 10 kpc, which is five scale lengths and therefore a very conservative estimate. Such a profile would have a mass of $1.5 \times 10^{10} M_{\odot}$, or just 5\% of the missing baryons from the Galaxy. 

As a consistency check, we compute the cooling time of this halo, using our halo temperature and metallicity with the cooling function of Sutherland and Dopita (1993). Our model assumes a stable halo, but we find the halo is probably not quite stable against cooling over a Hubble time. For $n=5\times10^{-4}$ cm$^{-3}$ (the upper limit on the average gas density out to the LMC), the cooling time is $\tau \approx 10$ Gyr. For the above NFW profile, then, $6.4\times10^9 M_{\odot}$ (40\% of the total hot halo gas mass) will cool within the age of the Galaxy and re-accrete onto the disk. To maintain self-consistency, the constraints above must be treated as upper limits on an even more rarefied halo, or there must be continuous re-heating of the halo gas by a wind or some other mechanism. 

\subsection{Other Constraints}

We can compare the constraints on the Galactic hot halo from LMC pulsars to other constraints derived using a variety of methods. For reference, the DM upper limit on the electron density at $r = 50$ kpc is  $n_e \le 7.7\times10^{-5}$ cm$^{-3}$ for an NFW halo, or $n_e \le 1.0 \times10^{-4}$ cm$^{-3}$ if we truncate the profile at $r_{\text{in}} = 10$ kpc. 

Our first comparison is to the result of Stanimirovi\'{c} et al. (2002), who used 21 cm observations of the Magellanic Streams to measure the radius and velocity width of clumps of neutral gas in the streams. From these measurements, they derive the internal pressure of the clumps, and equate this to the gas pressure of the hot halo (making the assumption that the clumps are confined entirely by gas pressure, and neglecting magnetic and turbulent pressure; this is therefore probably an upper limit on the gas pressure). They assume a halo temperature of $1 \times 10^6$ K; for our higher temperature of $3.5 \times 10^6$ K, the inferred upper limit on the hot halo electron density is $n$(45 kpc)$\le 9\times10^{-5}$ cm$^{-3}$ - consistent with the above constraint. 

We also consider high velocity clouds, using UV measurements of the temperature and density of numerous partially-ionized metals and inferred gas pressures (Fox et al. 2005). This analysis suffers from a size-distance degeneracy, however, and so Fox et al. perform their analysis for several potential cloud distances. The lower distance (10 kpc) yields a lower upper limit, and the higher distance (100 kpc) yields a higher upper limit, but they seem to prefer a distance of 50 kpc based on a tentative association of some of these clouds with the Magellanic Stream. Using this distance and converting to our halo temperature, the upper limits on the electron density at this radius are approximately $n_e = 3-6 \times 10^{-5}$ cm$^{-3}$. 

We also estimate density using the emission measure of the hot halo, as measured by Kuntz and Snowden (2000). They decompose the ROSAT soft X-ray background into components, tentatively identifying the harder emission from an unidentified component as the hot halo of missing baryons. This emission has a temperature of about $3\times10^6$ K, close to our estimate of the hot halo temperature, and it  appears to have a fairly uniform distribution across the sky. There is also a softer background component with a temperature of $T \sim 1\times10^6$ K, but this temperature is too cool to remain in a hot halo out to the virial radius, and its spatial distribution is not isotropic: it follows the Galactic disk.
 
The hotter component of the Galactic X-ray background has a best-fit emission measure of $2.1 \times 10^{-3}$ cm$^{-6}$ pc, for an assumed solar-metallicity gas. This corresponds to an NFW hot halo of total mass $3.1 \times10^9 M_{\odot}$ (1.0\% of the missing baryons from the Galaxy), or a truncated NFW halo of mass $4.1\times10^9 M_{\odot}$ (1.4\% of the missing baryons). Using our assumed metallicity of $0.2 Z_{\odot}$, we recalculated the emissivity using the XSPEC utility with an APEC model (Smith et al. 2001) and all other parameters matched to the observations in Kuntz and Snowden (2000), which makes these halo masses increase by 190\%. The corrected halo masses match very closely with the previous constraints: $5.9 \times 10^9 M_{\odot}$ (2.0\% of the missing baryons) for an NFW halo, $7.8\times10^9 M_{\odot}$ (2.6\% of the missing baryons) for a truncated NFW halo. This halo has a cooling time larger than 10 Gyr for $r > 15$ kpc (no truncation) or $r > 18$ kpc (with truncation).

We summarize all of these constraints in Table 1. Upper limits on the total hot halo mass, electron density at a radius of 50 kpc, and fraction of the missing baryons from the Galaxy in the halo, are computed for each constraint. The ranges on some parameters reflect ranges in the original constraint combined with fits to both a truncated and a non-truncated NFW profile. We also present the upper limits in terms of a flattened, power-law profile, discussed further in section 7. 

\begin{sidewaystable}[h]
\caption{Constraints on the Milky Way Hot Halo}
\begin{tabular}{|ccccccccc|}
\hline
  & \multicolumn{3}{c}{NFW profile} & & \multicolumn{3}{c}{Flattened profile} &\\
 \cline{2-4} \cline{6-8}\\
Method & \footnotesize{Hot halo mass} & $n_e$(50 kpc) & frac & & \footnotesize{Hot halo mass} & $n_e$(50 kpc) & frac & Reference\\
 & ($10^9 M_{\odot}$) & ($10^{-5}$ cm$^{-3}$) & & & ($10^9 M_{\odot}$) & ($10^{-5}$ cm$^{-3}$) & &\\
\hline \hline
\footnotesize{LMC Pulsar DM} & $<12-15$ & $<7.7-10$ & $<0.04-0.05$ & &$<170$ & $<27$ & $<0.58$ & \S 3.1\\ 
\footnotesize{Mag. Stream HI} &$<10-11$ & $<7$ & $<0.03-0.04$ & & $<53$& $<8$ & $<0.18$ & \footnotesize{Stanimirovi\'{c} et al. (2002)}\\
\footnotesize{HVCs} & $<5-9$ & $<3-6$ & $<0.02-0.03$ & & $<19-39$ & $<3-6$ & $<0.06-0.13$ & \footnotesize{Fox et al. (2005)}\\
\footnotesize{Galactic XRB} & $<5.9-7.8$ & $<4-5$ & $<0.02-0.03$ & & $<110$ & $<17$ & $<0.38$& \footnotesize{Kuntz \& Snowden (2000)}\\
\hline
\end{tabular}\\
\scriptsize{Parameters determined from fitting profiles to the constraints noted in section 3.2. {\it frac} is the ratio of the mass of the hot halo to the mass of the missing baryons from the Galaxy ($3\times10^{11} M_{\odot}$). }
\end{sidewaystable}

Finally, we also include a lower limit on the Galactic hot halo in addition to the upper limits previously noted. Blitz and Robishaw (2000) measure HI mass for the dwarf spheroidal galaxies in the Local Group, and use these masses to place upper limits on the amount of ram-pressure stripping that these galaxies could have experiences. This leads to an upper limit on the density of the Local Group intragroup medium, corresponding to $n_e \ge 2.5 \times 10^{-5}$ cm$^{-3}$. The Galactic hot halo must be denser then the ambient medium around it, so this figure also serves as a strict lower limit on the density of the hot halo. 

Our picture of the Milky Way hot halo is therefore a halo obeying an NFW profile, and containing several $\times$  $10^9 M_{\odot}$ of material. It extends out to at least 50 kpc, although its density declines to the IGM density by a radius of 80 kpc or so. Such a halo is fully consistent with all of the constraints we have identified, but it does not contain more than a few percent of the missing baryons from the Galaxy.

\section{OVII Absorption in Other Galaxies}

The previous techniques will not work for constraining the haloes of most other galaxies, since we cannot measure the DM of distant extragalactic pulsars, and even if we could, we cannot reliably separate the dispersion from intergalactic electrons from the dispersion caused by the galaxy's halo of missing baryons. However, the integrated column density of free electrons from the hot haloes still will have a measurable effect in other galaxies through OVII absorption features along quasar sightlines. 

At the temperatures of the haloes around large galaxies (a few hundred eV), intergalactic Oxygen exists in the highly ionized OVII or OVIII states, of which the former can be detected through its 21.60 $\AA$ K$\alpha$ absorption line if superimposed in front of a bright background X-ray source. Zero-redshift OVII absorption has been detected several times, and attributed variously to the Local Group IGM (Nicastro et al. 2002, Rasmussen, Kahn, and Paerels 2003) or to a hot halo around the Milky Way (Wang et al. 2005, Fang et al. 2006, Bregman and Lloyd-Davies 2007). In addition to the local detections, there are a few published claims of nonlocal OVII absorption lines (Mathur, Weinberg, and Chen 2003, Nicastro et al. 2005b, Buote et al. 2009), but these have yet to yield any significant constraints (see, e.g., Kaastra et al. 2006).

One successful approach comes from a statistical argument by Fang et al. (2006). They studied 20 active galactic nucleus (AGN) sightlines without any detections of nonlocal OVII absorption and argued based on the number density of L$_*$ galaxies and of Local Group-sized galaxy groups that most of the missing baryons must be restricted to haloes around the former; if these haloes extended to megaparsec scales, it would be statistically unlikely ($\sim 3.7 \sigma$) not to have seen them in OVII absorption. 

We extend this result by examining observations along specific sightlines, and broadening the search to galaxies smaller than $L_*$. We select critical sightlines with the best combination of redshift interval and high signal-to-noise observations discussed by Fang et al. (2006), and count the intervening galaxies along the sightline, focusing on the twelve AGN that fall within the coverage of the Sloan Digitial Sky Survey (SDSS) Seventh Data Release (Abazajian et al. 2009). For each of these twelve sightlines, we searched the SDSS archive for foreground galaxies whose virial radii overlapped the sightline. If such a galaxy is present, but no OVII absorption is detected at its redshift, then we can place an upper limit on the OVII column from the galaxy.

We searched within a $4^{\circ} \times 4^{\circ}$ region around each AGN for SDSS galaxies with photometric redshifts less than the AGN's redshift. We also found a known Lyman-$\alpha$ absorption line at each of these photometric redshifts to help verify the photometric redshifts; see Table 2. The photometric redshift was used to compute the distance to each galaxy and the angular distance between each galaxy and the AGN sightline. The virial radius for each galaxy was estimated, and galaxies were selected whose virial radii were larger than their distance to the AGN sightline. We inferred the virial radius by computing the SDSS g-band galaxy luminosity and assuming a scaling relation between luminosity and virial radius, normalized to the Milky Way ($L_g = 1\times10^{10} L_{\odot}$, $R_{\text{vir}} = 250$ kpc). We consider two forms for the scaling relation. The first, (scaling relation A), is based on empirical scaling between the baryon fraction (taken to be a proxy for luminosity) and the total mass of a galaxy (McGaugh 2000, Dai et al. 2010): $f_b \propto L/M  \propto M^{0.5}$, so $L \propto M^{3/2}$ and $R \propto L^{2/9}$. This seems to represent the data over many orders of magnitude in galaxies with disk components (McGaugh et al. 2005). For elliptical galaxies, the well-established "tilt" in the fundamental plane (e.g. Tortora et al. 2009) has the baryon fraction scaling in the opposite direction with mass: $M/L \propto L_B^{1/4}$, so $L \propto M^{4/5}$ and $R \propto L^{5/12}$ (scaling relation B). We classify each galaxy as early-type or late-type using its SDSS $u - r$ color, with early-type galaxies having $u-r > 2.2$ and late-type galaxies having $u-r < 2.2$ (Strateva et al. 2001, Mateus et al. 2006). For the former, we apply scaling relation B, and for the latter we apply scaling relation A to estimate the virial radius.

Using this technique, out of our sample of twelve sightlines we find three with nearby galaxies whose virial radii encompass the sightline:  3C273 ($z_{em} = 0.158$), Q1821+643 ($z_{em} = 0.297$), and Ton 1388 ($z_{em} = 0.177$). For each intervening galaxy, the mass and the baryon fraction are estimated, as well as  the expected mass in missing baryons. We assume these baryons are distributed in a halo around the galaxy, extending to the virial radius, and that they obey an NFW density profile (using several values for the concentration parameter between c=9 and c=17, as in Klypin, Zhao, and Somerville 2002).  The expected electron column contributed by each galaxy is computed by integrating along the AGN sightline as it passes through the halo. This gives us $N_{e, NFW}$, the electron column expected from a hot halo in hydrostatic equilibrium around each galaxy. 

In Table 2, we compare this column to the minimum detectable electron column inferred from Fang et al. (2006). For each sightline, they provide the expected width of a $3\sigma$ detection of OVII absorption, from which we compute the OVII column that would produce this line using N(O VII) = $3.48\times10^{14}$ EW, where N is in cm$^{-2}$ and EW in m$\AA$ (Bregman and Lloyd-Davies 2007). The electron column is estimated from the Oxygen column by adapting the conversion derived by Bregman and Lloyd-Davies (2007):\\

\[N_e = 2.9\times10^{20} \left( \frac{N_{O VII}}{10^{16} \text{ cm}^{-2}}\right) \left(\frac{f}{0.5}\right)^{-1} \left(\frac{Z}{0.2 Z_{\odot}}\right)^{-1}\text{ cm}^{-2}\]

where $f$ is the O VII ion fraction (we use $f = 0.5$) and assuming gas of $Z = 0.2 Z_{\odot}$. This yields $N_{e,th}$, the minimum (threshold) electron column that can be detected at $3\sigma$ using OVII absorption. These assumptions are conservative. The fraction of Oxygen in the O VII state is close to unity for temperatures between $3.5 \times 10^5$ K and $1.5 \times 10^6$ K, which are the expected temperatures for hydrostatic haloes around most of our selected galaxies. This is probably a low estimate since hot intracluster medium gas typically has a metallicity of about $Z = 0.2 Z_{\odot}$, even towards the virial radius of distant galaxy clusters (Maughan et al. 2008, Anderson et al. 2009). For both $f$ and Z, choosing less conservative values reduces $N_{e, th}$, which would imply even stronger expected detections than those listed in Table 2.

\begin{sidewaystable}[h]
\caption{OVII Sightline Analysis}
\begin{tabular}{|ccccccccccc|}
\hline
Sightline  & photo-z& \footnotesize{Ly$\alpha$ EW} & b & L$_g$ & u - r & N$_{e, th}$ & N$_{e, NFW}$ &(S/N)$_{\text{NFW}}$ & N$_{e,pl}$ & (S/N)$_{\text{pl}}$\\
&  & (m$\AA$)& \footnotesize{(kpc)} & (L$_{\odot}$)& &\scriptsize{$(10^{19}$ cm$^{-2}$)} &  \scriptsize{($10^{19}$ cm$^{-2}$)} & &\scriptsize{($10^{19}$ cm$^{-2}$)} & \\
\hline \hline
3C273 & $0.11\pm0.04$& $138^{\scriptsize{\text{a}}}$& 41 & $9\times10^8$ & 2.00 & $9.1$& $13.8^{+20.4}_{-10.2}$&4.6&$13.8^{+22.3}_{-12.2}$&4.5\\
Ton 1388 &$0.09\pm0.02$ &121$^{\scriptsize{\text{b}}}$ & 28 &$9\times10^7$ &2.77  &$12.0 $&$0.2^{+1.3}_{-0.2}$&0.1&$0.7^{+2.8}_{-0.7}$&0.2\\
Ton 1388 &$0.17\pm0.01$ &233$^{\scriptsize{\text{b}}}$ &39&$4\times10^9$ &3.21  &$12.0 $ &$20.5^{+12.2}_{-8.4}$&5.1&$16.9^{+23.2}_{-13.7}$&4.2\\
Ton 1388 &$0.06\pm0.03$ & 79$^{\scriptsize{\text{b}}}$&43& $5\times10^8$&1.76  & $12.0$& $9.2^{+17.9}_{-7.8}$&2.3 &$11.1^{+19.6}_{-10.7}$&2.8\\
Ton 1388 &$0.13\pm0.01$ & 132$^{\scriptsize{\text{b}}}$&135&$1.2\times10^{10}$ &3.07 & $12.0$&$6.2^{+3.5}_{-2.5}$&1.5&$11.8^{+16.1}_{-6.1}$&2.9\\
Q1821+643 & $0.21\pm0.05$ &560$^{\scriptsize{\text{c}}}$ &75  &$1\times10^9$ & 3.40 &$10.0$&$0.9^{+3.0}_{-0.9}$&0.3 &$2.7^{+7.1}_{-2.7}$&0.8\\
Q1821+643 & $0.18\pm0.10$ &350$^{\scriptsize{\text{c}}}$& 96 & $4\times10^9$&1.81 &$10.0$ &$6.4^{+12.3}_{-5.6}$&1.9&$11.2^{+19.9}_{-7.3}$&3.4\\
\hline
\end{tabular}
\scriptsize{The impact parameter $b$ is the distance between the galaxy and the relevant AGN sightline. Each galaxy was matched to Lyman $\alpha$ absorption detected at or very near its photometric redshift; the equivalent with is noted. The neutral Hydrogen column densities implied by the Ly-$\alpha$ lines are $\sim 10^{13} - 10^{14}$ cm$^{-2}$ - significantly lower than the electron column densities, as would be expected from highly ionized plasma. The Lyman $\alpha$ references are: (a) Sembach et al. 2001, (b) Sembach et al. 2004, (c) Bahcall et al. 1993. The $u - r$ color is used to distinguish between early-type ($u - r > 2.2$) and late-type ($u - r < 2.2$) galaxies when computing $N_{e, NFW}$. EW$_{th}$ is the equivalent width of a $3\sigma$ OVII detection for the {\it Chandra} observation of this AGN in Fang et al. (2006), and N$_{e, th}$ is the electron column corresponding to this EW. N$_{e, NFW}$ is the electron column we expect from the hot halo around this galaxy if it contains the missing baryons (the uncertainty reflects $1\sigma$ uncertainties propagated from the redshift, an assumed 5 kpc uncertainty in the impact parameter, the g-band photometry, scaling relation, and different choices for C in the NFW profile).  (S/N)$_{\text{NFW}}$ is the strength of the detection we would expect for this halo if it extended to the virial radius. N$_{e,pl}$ and SN$_{\text{pl}}$ are the column density and signal-to-noise of a flatter, power-law profile that will be discussed in section 7.}
\end{sidewaystable}

Table 2 shows that two of these galaxies should have been detected in existing observations. The nondetection of these absorption lines implies real electron column must be much smaller than the N$_{e, NFW}$ value derived above. In light of our conservative assumptions about unknown parameters like $f$ and $Z$, the most reasonable way to reduce $N_{e, NFW}$ below $N_{e, th}$ is to bring down the mass of the haloes. The radius could also be increased to several times the virial radius, which would reduce the density but increase the path length, but this would also increase the number of galaxies whose haloes intersect with the AGN sightlines. For example, if R is doubled to twice the virial radius, we would expect 4$\times$ as many absorbers, yielding about 6 detections of 3$\sigma$ or more. Reducing the halo radius below the virial radius increases the density, which worsens the pulsar dispersion measure inconsistencies in the previous section and leads to a problem of excess X-ray emission (see next section).  Also, our strongest constraints come from two galaxies of 5-10\% of $L_*$, each within about 40 kpc of the sightline. As long as these galaxies have haloes at least 40 kpc in extent, we can put a $3\sigma$ upper limit on the mass in the halo of 70\% of the missing baryons, so reducing the radius over a reasonable range will not have much of an effect. Thus the only remaining option is to lower the mass, below 70\% of the missing baryons at 3$\sigma$.

This is a stronger conclusion than Fang et al. (2006) reached in their analysis. They were able to rule out  haloes with columns greater than N$_{e, th}$ with radii of 1 Mpc at $3.7 \sigma$, but they did not rule out such haloes with R $\sim$ 100 kpc. Their statistical analysis predicted only 0-1 $L_*$ galaxies to fall within 100 kpc of a sightline in their entire sample, so their method did not really constrain the parameters of haloes of this size. Indeed, we find only one $L_*$ galaxy, and it lies at an impact parameter of 135 kpc, in agreement with their prediction. However, there are several $0.1 L_*$ galaxies closer than 100 kpc, and these are critical in deriving our constraints.  At larger radii, our predictions are also consistent with not observing excess OVII  absorption in a sightline that passes within 350 kpc of M31 (Bregman and Lloyd-Davies 2007). 

\section{X-Ray Surface Brightness}

The failure to detect extended hot gaseous halos from X-ray emission (e.g. Strickland et al. 2004, Li et al. 2006, Li, Wang, and Hameed 2007) places particularly strong constraints on the fraction of missing baryons present around galaxies. Unlike the other constraints in our paper, this constraint has been known for at least a decade; see, for example, Benson et al. (2000) and Wu, Fabian, and Nulsen (2001). The former paper worked out X-ray luminosities of hot haloes around massive late-type galaxies, and found these should be some of the most luminous X-ray sources in the local Universe. The latter paper explored the unresolved X-ray background, and noted that the predicted contribution to the background from hot haloes around galaxy groups was an order of magnitude larger than the unresolved background. Since these haloes are so large and diffuse, however, measurements of both the luminosity and the unresolved background number density require extrapolation out to large radii and low surface brightness. Therefore, we instead consider this discrepancy in terms of X-ray surface brightness, since this quantity is directly measurable and can easily be used to establish observational constraints on these hot haloes. 

We calculate the X-ray emissivity for three different galaxy masses: $3\times10^{11} M_{\odot}$, $8 \times10^{11} M_{\odot}$, and $2 \times 10^{12} M_{\odot}$. For each mass, the hot halo is assumed to obey an NFW profile with a concentration C=12 (the profile is not very sensitive to this value), and the hydrostatic equilibrium temperature is computed. The emission from the baryons in the hot halo is modeled with an optically thin emission model (the APEC model) with photoelectric absorption from Galactic neutral Hydrogen of column density N$_H = 3\times10^{20}$ cm$^{-2}$. For each halo, we use the WebSpec utility to compute the flux $F_1$ over an energy range of 0.1-6 keV on the ACIS-S instrument. The metallicity was set to $0.2 Z_{\odot}$. 

We estimate a typical Chandra ACIS background of $1.3\times10^{-12}$ erg s$^{-1}$ arcmin$^{-2}$ from, e.g., Bregman et al. (2009), which is converted into a flux (in erg s$^{-1}$ cm$^{-2}$) for each halo by dividing by the telescope effective area A determined using XSPEC. Assuming halo emission can be detected once it reaches a surface brightness of about 30\% of the background, this background can be combined with the flux $F_1$ from above and with the definition of normalization in the APEC model (at $z=0$, norm $= n^2 R\times10^{-14} / 4\pi$) to calculate the minimum detectable emission measure for gas in each halo:

\[(n^2R)_{\text{min}} = \frac{4 \pi \times 0.3 \times (1.3 \times 10^{-12} \text{ erg s}^{-1} \text{ arcmin}^{-2} )}{10^{-14} \times  F_1 \times (8.5\times10^{-8} \text{ arcmin sterad}^{-1}) \times A} \text{ cm}^{-5}\]

The results of this analysis are presented in Table 3.

\begin{table}[h]
\caption{Halo Emissivity Analysis}
\begin{tabular}{|ccccc|}
\hline
Mass  & $kT$& $F_1$ & A & $(n^2R)_{\text{min}}$  \\
($M_{\odot})$ & (keV) & (erg s$^{-1}$ cm$^{-2})$ & (cm$^{2})$ &(cm$^{-6}$ kpc)\\
\hline \hline
$2\times10^{12}$ & 300 & $4.06 \times 10^{-10}$ & 217 & $2.1\times10^{-5}$\\
$8\times10^{11}$ & 160 & $2.31\times10^{-10}$ & 137 & $5.9\times10^{-5}$\\
$3\times10^{11}$ & 85  & $5.12\times10^{-11}$ & 62& $5.9 \times 10^{-4}$\\
\hline
\end{tabular}\\
\scriptsize{Numbers used for derivation of emission measure ($n^2R)_{\text{min}}$ of halo gas in hydrostatic equilibrium that would be detected by {\it Chandra} ACIS-S at 30\% of background. Mass is the total mass of the galaxy and dark matter halo, and $kT$ is the temperature of the halo gas. $F_1$ is the estimated flux from an APEC model (see text). And A is the effective area of the detector for each observation.}
\end{table}

We integrate along columns through the halo to find the expected EM at each projected radius from this galaxy. In Figure 1, we compare the expected emission measure to the minimum detectable EM for galaxies of various concentrations at all three masses. While the emission within the disk may be obscured by or confused with the disk, the haloes can be detected out to many times the radius of the thick disk in all three cases. A Milky Way-sized halo, for example, should be detectable in emission out to about 75 kpc, and even a hot halo around a galaxy of total mass $3 \times 10^{11} M_{\odot}$ should be detectable out to twice the radius of the thick disk. 

These haloes should be very obvious in X-rays around all reasonably-sized galaxies, and yet the observed X-ray haloes are confined to the immediate vicinity of the thick disk. This is a serious problem for the paradigm of hot haloes full of missing baryons. The missing baryon electron density must be smaller than the model predicts; the upper limit on the mass of the halo to bring the emissivity at the edge of the thick disk down to the detection threshhold is between 11\% and 24\% for the three different masses. In general larger galaxies have stricter upper limits, since they have more mass in their haloes. Therefore, this constraint can be improved further using deep X-ray observations of very massive isolated disk galaxies. 

Observations can also constrain the density profile of the hot halo; based on the predicted EM, in the detectable region outside the thick disk our NFW profiles roughly obey a $\beta$ model with $\beta \approx 0.87$. This is slightly steeper than the profiles of hot gas in early-type galaxies, which have $\beta \approx 0.4-0.6$ (e.g. Brown and Bregman 2001). In Figure 5, we also consider a flatter density profile as discussed further in section 7. This profile has $\beta \approx 0.36$ in the observable region, shallower than observed gas profiles. However, since observed values of $\beta$ fall in between these two profiles, it seems likely that real haloes are neither as steep as NFW profiles nor as shallow as the flattened profiles of section 7.

\section{Energetics for Gas-Rich Late-Type Galaxies}

Having established some observational constraints on hot haloes around the Milky Way and other large galaxies, we comment briefly on the overall energetics of expelling the missing baryons from galaxies (either into virialized haloes or outside of the entire system) using large-scale galactic winds. These winds are a common component of the current generation of semi-analytic models of galaxy formation, required to reconcile the dominant ``cold flows'' picture (e.g. Kere\u{s} et al. 2005), which assumes galaxies accrete the cosmological fraction of baryons, with the observed severe baryon depletion in nearly every galaxy today. Winds are also observed directly at $z \sim 2-3$ in galaxies undergoing periods of intense star formation and AGN activity (e.g. Pettini et al. 2001, Geach et al. 2005, Law et 
al. 2007), but appear insufficient to expel the missing baryons (Martin 2005). 

Such winds come in two varieties: ``energy-driven'' and ``momentum-driven''.  The former carry kinetic energy proportional to the energy injected into the interstellar medium by supernovae and/or an AGN, while the latter (Murray, Quataert, and Thompson 2005) carry momentum proportional to the momentum injected into the ISM by supernova shocks and radiation pressure from stars and AGN. In both cases, however, the constant of proportionality (the ``efficiency'' of the coupling) has to be large in order to drive enough of a wind to expel the missing baryons. For example, recent simulations by Dutton and van den Bosch (2009) require 25\% efficiency for energy-driven winds or 100\% efficiency for momentum-driven winds to expel the missing baryons, and they neglect cooling of the ejected baryons, which can result in the baryons re-accreting onto the disk in a galactic ``fountain'' and can therefore increase the required ejection energy by several times. Oppenheimer and Dav\'{e} (2008) use momentum-driven winds with an equivalent efficiency of 200\% with respect to the energy available to energy-driven winds. Such high efficiencies are an attractive feature of momentum-driven winds, but we note that the theory still needs to be fully developed and observationally supported. In particular, it is not yet clear that momentum-driven winds are stable enough to drive most of the baryons from a galaxy instead of fragmenting into a less efficient multiphase wind, although further research may illuminate this question. 

Regardless, while requiring such a large efficiency may already be problematic, driving these winds also leads to some challenging conclusions. There are two major sources of energy/momentum available for this purpose: supernovae and accretion onto a supermassive black hole (SMBH). In addition, one or both of these can be triggered by major mergers, but mergers primarily increase the amount of energy in these two reservoirs rather than adding a new component. (Mergers also do release gravitational energy, but we are not aware of any way to couple efficiently this energy to the baryons in galaxies.) The maximum energy or momentum available in a galaxy from supernovae should be proportional to the stellar mass of the galaxy, and the maximum energy/momentum available from the AGN should be proportional to the mass of the black hole. Since the black hole mass is proportional to the galactic bulge mass (Marconi and Hunt 2003), which is dominated by stars, this would seem to suggest that the total energy available to eject baryons is related to the stellar mass of the galaxy, which would lead naturally to a baryonic Tully-Fisher relation (McGaugh 2005). 

However, this picture also leads to the prediction that variations in the energy sources should affect the baryon fractions of galaxies, and this is not observed. The observations show that, at least for disk galaxies, the baryon fraction seems to depend almost solely on the circular velocity at large radius (a proxy for total mass; McGaugh 2005, Stark, McGaugh, and Swaters 2009). But if the baryon fractions are caused by supernova- and AGN-driven winds, then galaxies with fewer supernovae or smaller SMBHs should have higher baryon fractions than galaxies with more supernovae or larger SMBHs. So, for example, galaxies with no central black hole such as M33 (Gebhardt et al. 2001) should have higher baryon fractions than galaxies with the the same total mass but a large SMBH, when in reality M33 lies on the BTF. 

If the supermassive black hole (SMBH) in a galaxy is responsible for a significant amount of the baryon loss, there should be a relationship between the baryon fraction and the galaxy type.  Equivalently, in a fit to the baryon Tully-Fisher relationship, galaxies with relatively larger SMBHs (earlier-type galaxies) would lie below the normal scaling relation if AGN activity from a SMBH expels gas.  To search for this effect, we examine the sample of galaxies from McGaugh (2005) and use the galaxy type as a proxy for the relative importance of a SMBH. When the galaxy type was not determined to sufficient precision, the galaxy was not used for this comparison.  For the 60 galaxies with good galaxy types, there is no trend of the residual with galaxy type (Figure 2). We estimate that we could have detected a trend of about 20\% or larger in the residual between Sb and Sd, suggesting AGN activity is responsible for removing a mass equivalent to less than 20\% of  the remaining baryons. Since present-day baryon fractions are typically 5-30\% of the cosmological value, this corresponds to AGN removing no more than 1-5\% of the original baryonic content from galaxies over their lifetime. 

We apply a similar analysis to examine the effect of supernova energy on baryon fraction. We computed the ratio of stellar mass to total baryon mass in the sample of gas-rich disk galaxies of Stark, McGaugh, and Swaters (2009). To estimate the total baryon mass, we assumed an NFW halo potential and applied the relation $M_{200} = 2.3\times10^5 v_f^3 h^{-1} M_{\odot}$ for $v_f$ in km s$^{-1}$ from Navarro (1998). The stellar mass is computed in Stark et al. for each galaxy from stellar population modeling. In Figure 3, we compare this ratio to the circular velocity at large radius $v_f$ for each galaxy. The figure shows these galaxies are all significantly depleted in stellar fraction compared to average disk galaxies, so there should be significantly less supernova energy available to remove baryons in these galaxies. We examined whether the residuals about the BTF are correlated with the degree to which the galaxies are gas-dominated.  No such correlation was found, using both parametric and non-parametric tests, which suggests that supernovae from stars in these galaxies do not appear to be responsible for removing most of the baryons. 

An alternative explanation of the lack of BTF correlation with galaxy type or gas fraction can be inferred from the result of Navarro and Steinmetz (2000). They noted that the remarkably low scatter in the I-band Tully-Fisher relation can be explained through galaxies with different disk mass fractions scattering along the Tully-Fisher relation instead of perpendicular to the curve. The idea is that the disk mass actually contributes to the measured rotation speed in addition to the halo mass, so if the disk mass increases, the rotation speed will also increase, regardless of the halo mass. This explanation does not apply directly to our argument, however. Neither the gas fraction nor the galaxy type correlates with the disk mass, so there is no reason to expect the scatter between these quantities and the rotation speed to follow the BTF. However, if supernovae or AGN accretion has expelled a dynamically significant mass of baryons from one of these galaxies, we would expect the rotation speed of the galaxy to decrease near the disk, in an analogous effect to the result of Navarro and Steinmetz. To minimize this concern, the rotation speeds used in our analysis are inferred from 21 cm emission at large radii, outside of the disk, where the rotation curve has flattened (for details, see McGaugh 2005 and Stark et al. 2009). At these radii (tens of kpc), the dark matter dominates the gravitational potential, and so the efficiency of expelling baryons should not affect the measured rotation velocity. Thus we do not believe correlations between baryon fraction and rotation velocity affect our claim that AGN accretion and supernovae are not primarily responsible for expelling the missing baryons. 

A simple way to understand the argument of this section is to consider two galaxies: NGC 4183 and NGC 4138. Figure 4 shows images of these two galaxies taken from the Sloan Digital Sky Survey, and in Table 4 we present measured or inferred parameters for the baryons in these galaxies, based on the results in NED and in Stark, McGaugh, and Swaters (2009) as described above. 

\begin{table}[h]
\caption{Comparison of Two Nearby Disk Galaxies}
\begin{tabular}{|ccccccccc|}
\hline
Name  &  i &  $v_f$ & $M_{200}$ & $M_b$ & $f_b$ & $M_g$ & $M_s$ & Type \\
 & \scriptsize{(degrees)} & \scriptsize{(km s$^{-1})$} & \scriptsize{(log $M_{\odot}$)} & \scriptsize{(log $M_{\odot}$)} & \scriptsize{($M_{b} / M_{200}$)}& \scriptsize{(log $M_{\odot}$)} & \scriptsize{(log $M_{\odot}$)} & \\
\hline \hline
NGC 4138  & 53 & $148 \pm 4$ & 12.02 &  10.34 & 0.021 & 9.30 & 10.30 & S0\\
NGC 4183 & 82 & $111\pm 2$ & 11.64 &  9.85 & 0.016 & 9.69 & 9.33 & SAcd\\
\hline
\end{tabular}
\scriptsize{\\Data taken from Stark, McGaugh, and Swaters (2009) and from the NASA Extragalactic Database, as described above. \\$M_{200}$ is computed from $v_f$ as described above as well, assuming $h = 0.72$. }
\end{table}

These two galaxies are similar in mass, but very different in composition. NGC 4138 is a bulge-dominated lenticular galaxy with 91\% of its baryons in stars. NGC 4183 is a late-type disk galaxy with only 31\% of its baryons in stars (the other 69\% are in gas). Our sample contains more extreme cases of both gas-rich and gas-poor galaxies than these, but these two present a useful contrast in both gas fraction and galaxy type, with little difference in total mass. Since NGC 4138 has such a high stellar fraction and such a prominent bulge, it must have been able to use its AGN and its supernovae to produce a significant galactic wind and drive out more of its baryons than NGC 4183, and yet the latter has a smaller baryon fraction than the former. 

More quantitatively, if we assume a negligible SMBH in NGC 4183 and assume its $2.1\times10^9 M_{\odot}$ of stars has produced $1.8\times10^7$ supernovae\footnote{This is estimated by integrating over a Chabrier (2003) initial mass function, letting stars with $M > 8 M_{\odot}$ explode in a core-collapse supernova, and applying a 20\% correction (Mannucci et al. 2005) to include type Ia supernovae}, we can estimate the average mass each supernova needs to eject to account for the missing baryons. From eq. (1), we calculate NGC 4183 is missing $7.9\times10^{10} M_{\odot}$ of baryons, so each supernova needs to heat and eject $4400 M_{\odot}$ from the galaxy. It already requires $180\%$ of the supernova energy (i.e., $1.8\times10^{51}$ erg) to heat $4400 M_{\odot}$ to the virial temperature of NGC 4183, and a comparable amount of energy to drive this material up into a halo from the gravitational potential of the disk, for a total supernova energetic efficiency of 350\% (or more, if cooling is included). These extreme efficiencies may pose a problem even for momentum-driven winds, and this leads to our suggestion that the classical picture of AGN- and supernova-driven winds is incomplete for explaining the missing baryon problem.

\section{The Effect of Flatter Density Profiles}

There exists one significant modification to the standard picture of hot haloes that must be considered. Significant pre-heating of the baryons can lead to excess entropy in the halo gas, which causes it to deviate from the dark matter NFW profile and obey a flatter density profile instead (e.g. Mo and Mao 2002, Kaufmann et al. 2009). We approximate the relaxed high-entropy profile in Figure 6 of Kaufmann et al. (2009) with a power-law:

\begin{equation} \rho(r) = \rho_0 \left(\frac{r}{r_{\text{vir}}}\right)^{-p}\end{equation}

\noindent with $p = 0.9$. We recomputed the Galactic constraints for this flattened density profile (see Table 1). The DM constraint is significantly relaxed: the halo can hold 58\% of the missing baryons before reaching the LMC DM limit of 23 cm$^{-3}$ pc. This limit is dramatically weaker compared to the NFW upper limit of about 5\%. On the other hand, the pressure confinement constraints based on high velocity clouds and Magellanic Stream clouds place an upper limit of $6-13\%$ of the missing baryons ($2-4\times10^{10} M_{\odot}$), which still fails to resolve the second missing baryon problem. A flattened profile containing all the missing baryons also violates the EM measurement of Kuntz and Snowden (2000). They measured an EM of $2.1 \times 10^{-3}$ cm$^{-6}$ pc (which becomes approximately $7.7 \times 10^{-3}$ cm$^{-6}$ pc for our assumed metallicity), while the flattened halo would have an EM of $1.9\times10^{-2}$ cm$^{-6}$ pc. Thus the picture we described in Section 3.2 remains - the Galaxy has a limited hot halo of some form that declines to the IGM density, but it does not contain most of the missing baryons from the Galaxy.

The extragalactic emission measure constraint (X-ray surface brightness) is also significantly affected by a flatter profile. Kaufmann et al. claim their profile lowers the X-ray luminosity of the inner region of the halo by a factor of $\sim 70$. We computed the emission measure for the flattened power-law profile (Figure 5) and indeed found the EM lowered by a factor of 70-300 at the edge of the thick disk. However, the emissivity of the gas per unit mass increases, by about 75\%, because the flattened profile is hotter than the NFW profile. Finally, since the slope of the profile is flatter, the emission can remain significant out to larger radii. The result of these combined effects is that the emission from within the disk is lowered significantly, but the emission remains flatter, so at larger radii it is still visible. As Figure 5 shows, smaller galaxy haloes are not detected, but a Milky Way-sized galaxy would still have a halo detectable out to 70 kpc or so, and a larger galaxy would have a hot halo visible to even larger radii. Searching for X-ray emission at significant fractions of the virial radius around large, isolated disk galaxies is therefore a good way to constrain these flattened profiles. 

This claim differs from the conclusions in Kauffmann et al. (2009). They argue that the surface brightness of their high-entropy halo is consistent with observations, while we argue the emission from such a halo should be detectable tens of kpc away from the galaxy disk. This distinction is primarily due to different assumptions about the baryon fraction. They initialize their models 10 Gyr ago with a baryon fraction $f_b = 0.1$, while we assume galaxies form with the cosmological baryon fraction $f_b = 0.17$. This gives our halo a density 1.7 times larger than theirs, which yields a factor of three in the emission measure. Correcting for this effect reconciles our respective predictions with each other, although we disagree on the interpretation. To understand the second missing baryon problem, however, one must begin with the cosmological baryon fraction, and then explain the fate of all the baryons originally associated with the galaxy over cosmic time; thus we argue that flattened density profiles do not hide all of the missing baryons from galaxies. 

The flattened profiles are still detectable through their OVII absorption as well. At large fractions of the virial radius, a flattened profile has a higher density compared to an NFW profile, so sightlines that only pass through a small section of the hot halo produce larger absorption lines if the profile is flatter. Sightlines that pass closer to the central galaxy produce roughly equivalent sightlines to the NFW halo. We computed the absorption lines expected for our seven absorbers, and displayed the results in Table 2 above. In the former regime, the small $(<1 \sigma)$ detections predicted for NFW haloes increased moderately if the profile was flatter and the $1-2\sigma$ detections became roughly $3\sigma$ detections. In the latter regime, the two large $>4\sigma$ detections both decreased slightly. Since a $3\sigma$ absorption line could be observable, the flattened density profiles do not seem to solve the OVII absorption issue, and actually make the problem worse.  

In general, flattened profiles can contain a larger fraction of the missing baryons from galaxies than NFW profiles. For the Milky Way, a flattened profile can meet the LMC pulsar DM constraint with up to 58\% of the missing baryons in the hot halo instead of 5\%, although the pressure constraints bring this figure down to 6-13\%. The X-ray emissivity argument constrains the mass of these haloes around other large galaxies to about 59\% of the missing baryons from these galaxies. This is a factor of three more mass than the NFW profile can hold without violating surface brightness limits.  The OVII argument also limits the mass of these haloes in the same way it limits the mass of NFW haloes, with an independent upper limit of about 70\% on the fraction of the missing baryons that can reside in the halo. Flattened haloes are therefore an improvement, but they do not seem to be a perfect solution.

If the halo is driven from the galactic disk by a superwind, then the energetics problem we highlighted still applies. The flattened profile puts some additional constraints on the wind, however. In Kaufmann et al. (2009), the high entropy that keeps the profile flat is assumed to be injected at redshift 2 or 3. This pushes back the energetics problem 10 Gyr or so, but still does not quite resolve it. On the other hand, if the halo has existed in its present form for 10 Gyr, perhaps the gas simply never fell into the disk in the first place. We will discuss this idea further in the conclusion.

\section{Discussion and Conclusion}

The above arguments place separate and independent constraints on the baryons around galaxies. These arguments lead us to two general conclusions:

First, the majority of the missing baryons from spiral galaxies do not lie in hot haloes around these galaxies. For the Milky Way, the combination of observed constraints suggests that the common assumption of a hot halo of missing baryons obeying an NFW density profile with a concentration of 12 can only hold 2-3\%  of the baryons missing from the Galaxy.  The emission measure of theoretical haloes places an upper limit of 11-24\% on the fraction of the missing baryons that can reside in similar hot haloes around other large galaxies.

The analysis of O VII sightlines constrains hot haloes around a wider mass range of galaxies. The free electrons in these haloes should be detected through O VII absorption in the spectra of several strong background X-ray sources. This absorption has not been observed, so the column must be several times smaller than the prediction. This would occur if the halo has a radius of $\lapprox 40$ kpc or if its mass is reduced by factor of 3-4. The former case leads to further difficulties.  The density would be 0.01-0.1 cm$^{-3}$, so the halo would cool quickly and should be far too bright in X-rays. The latter case is possible, but such a rarefied halo would not account for the missing baryons from galaxies. 

The proposal of hot haloes obeying flattened density profiles instead of NFW profiles was also considered. Around the Milky Way, a typical flattened profile could hold $6-13\%$ of the missing baryons without violating any of our constraints. The emission measure of hot haloes with a flattened profile around other large disk galaxies sets an upper limit on the mass of about 59\% of the missing baryons from these galaxies. The OVII constraint is approximately unchanged, which also limits the mass and extent of flattened density profiles.

Collectively, these arguments rule out the existence of a large reservoir of hot gas around large galaxies that could account for the missing baryons. The most likely scenario is the presence of a small gaseous halo around $L_*$ galaxies containing $10^9 - 10^{10} M_{\odot}$ and extending to a radius of 50-100 kpc. This picture also satisfies pulsar DM observations, measurements of the pressure around high-velocity clouds, and X-ray surface brightness constraints. If such a limited halo exists, it is not the primary reservoir of the missing baryons from galaxies.

Our second general conclusion is that galaxies do not expel their baryons primarily by galactic winds driven by supernovae or AGN activity.  We infer this from the lack of correlation between missing baryon fraction in galaxies and the stellar fraction or bulge mass in the galaxy. Since the latter observables are taken to be proxies for the energy available in a galaxy from supernovae or from a supermassive black hole, the lack of correlation implies these energy sources do not explain the observed missing baryon fractions. 

Additionally, there are galaxies where most of the baryons are still gaseous yet they have the same missing baryon fraction as their counterparts where the baryons are nearly all in stars. To explain the baryon depletion of some of these star-poor, gas-rich galaxies, the energy required is several times that expected from galactic winds powered by supernovae. 

It is beyond the scope of this paper to model the remaining mechanisms for resolving the second missing baryon problem, but we can comment on the possible solutions. Based on our conclusions, a straightforward way to account for the missing baryons from galaxies is if the baryons never fell into galaxies in the first place. This would require pre-heating before or just as the galaxy is forming. The most obvious source of non-gravitational energy at these times is early supernovae, whose contribution to heating would be much more effective in the epoch of protogalaxies (when potential wells are shallow) than it is today. These early supernovae could even come from the hypothetical ``Population III'' stars, which have the advantage of a top-heavy IMF, so this process would not convert much of the missing baryonic mass into low-mass stars.

If the missing baryons never accreted onto protogalaxies because of early supernova heating, there are several observable consequences in addition to resolution of the second missing baryon problem. These early supernovae pre-enrich gas in addition to pre-heating it, and this helps to explain phenomena like the G-dwarf problem and the "floor" in iron abundance noted in section 4. For more discussion of this possibility, see Benson and Madau (2003). Due to the top-heavy IMF, they would have short lifetimes, so few of these early unbound stars would survive today, and the strict limits on intracluster light (e.g Krick and Bernstein 2007) would not be violated. A top-heavy IMF would also increase the efficiency per unit mass of metal production, which has further implications for the metallicity in galaxy clusters, as discussed in Bregman, Anderson, and Dai (2010). There are also observable constraints on this process. For example, if supernova heating is to prevent the baryons from accreting onto galaxies, the supernovae must occur before the galaxies form, i.e. $z \gapprox 6$. There is also an important period at $z \sim 8$ when Compton cooling can efficiently cool the baryons (Cen 2003). Depending on the details of the supernovae and the cosmology, the bulk of the early star formation and baryon expulsion might occur either before or after this period. More detailed observation and theoretical work can further constrain these possibilities. 

The remarkably tight correlation for galaxies between dark matter halo mass and baryon fraction seems to offer a powerful clue for understanding galaxy formation. The correlation persists over variations in star/gas mass ratio  The mechanism and details of this connection remain unclear, however, and still need to be understood. Further observations could also help to constrain more tightly the mass and extent of hot haloes around galaxies. Direct imaging around large galaxies in soft X-rays could strengthen the surface brightness constraints. Measurements of extragalactic pulsars or additional OVII lines could also provide useful constraints on the electron column density around galaxies. Detailed examination of simulations could also be helpful, predicting, for example, the fraction of dark matter in the Universe that falls into galactic haloes for comparison to the behavior of the baryons. Finally, more modeling of galaxy formation could greatly assist in producing testable predictions that would differentiate between various solutions to this second missing baryon problem.

\section{Acknowledgements}
We would like to thank the anonymous referee for a thoughtful and helpful report that greatly improved this manuscript. We wish to thank M. Ruszkowski, S. McGaugh, O. Gnedin, E. Bell, and A. Muratov for helpful comments and discussions that related to this work. This research has made use of the Sloan Digitial Sky Survey. Funding for the SDSS and SDSS-II has been provided by the Alfred P. Sloan Foundation, the Participating Institutions, the National Science Foundation, the U.S. Department of Energy, the National Aeronautics and Space Administration, the Japanese Monbukagakusho, the Max Planck Society, and the Higher Education Funding Council for England. The SDSS Web Site is http://www.sdss.org/. This research has made use of NASA's Astrophysics Data System. This research has made use of the NASA/IPAC Extragalactic Database (NED) which is operated by the Jet Propulsion Laboratory, California Institute of Technology, under contract with the National Aeronautics and Space Administration. We gratefully acknowledge support from NASA in support of this research.  MEA also acknowledges support from the NSF in the form of a Graduate Research Fellowship.

\section{References}
\noindent Abazajian, K. N. et al. 2009, ApJS 182:543\\
Anders, E. and Grevesse, N. 1989 Geochim. Cosmochim. Acta 53:197\\
Anderson, M. E. et al. 2009, ApJ 698:317\\
Bahcall, J. N. et al. 1993, ApJS 87:1\\
Bell, E. F. et al. 2003, ApJ 585:117L\\ 
Bellazzini, M. 2004, MNRAS 347:119\\
Benson, A. J. et al. 2000, MNRAS 314:557\\
Benson, A. J. and Madau, P. 2003, MNRAS 344:835\\
Binney, J. and Tremaine, S. 2008, {\it Galactic Dynamics, 2nd Ed.} Princeton: Princeton U.P.\\
Blitz, L. and Robishaw, T. 2000, ApJ 541:675\\
Bouch\'{e}, N. et al. 2009, preprint (astro-ph/0912.1858)\\
Bower, R. G. et al. 2006, MNRAS 370:645\\
Bregman, J. N. 2007, ARA\&A 45:221\\
Bregman, J. N. and Lloyd-Davies, E. J. 2007, ApJ 669:990\\
Bregman, J. N. et al. 2009, ApJ 699:1765\\
Bregman, J. N., Anderson, M. E., and Dai, X. 2010, submitted to ApJ\\
Brown, B. A. and Bregman, J. N. 2001, ApJ 547:154\\
Buote, D. et al. 2009, ApJ 695:1351\\
Cen, R. and Ostriker, J. P. 1999, ApJ 514:1\\ 
Cen, R. 2003, ApJ 591:12\\
Cen, R. and Ostriker, J. P. 2006, ApJ 650:560\\
Chabrier, G. 2003, PASP 115:763\\
Croton, D. J. et al. 2006, MNRAS 365:11\\
Dai, X. et al. 2010, submitted to ApJ \\
Dav\'{e}, R. et al. 2001, ApJ 552:473\\
Dav\'{e}, R. 2009, preprint (astro-ph/0901.3149)\\
Dekel, A. and Birnboim, Y. 2006, MNRAS 368:2\\
Dunkley, J. et al. 2009, ApJS 180:306\\
Dutton, A. A. and van den Bosch, F. C. 2009, MNRAS 396:141\\
Fang, T. et al. 2006, ApJ 644:174\\
Fox, A. J. et al. 2005, ApJ 630:332\\
Fukugita, M., Hogan, C. J., and Peebles, P. J. E. 1998, ApJ 503:518\\
Fukugita, M. and Peebles, P. J. E. 2004, ApJ 616:643\\
Fukugita, M. and Peebles, P. J. E. 2006, ApJ 639:590\\
Gaensler, B. M. et al. 2008, PASA 25:184\\
Geach, J. E. et al. 2005, MNRAS 363:1398\\
Gebhardt, K. et al. 2001, AJ 122:2469\\
Howk, J. C. et al. 2002, ApJ 569:214\\
Kaastra, J. S. et al. 2006, ApJ 652:189\\
Kaufmann, T. et al. 2009, MNRAS 396:191\\
Kere\u{s}, D. et al. 2005, MNRAS 363:2\\
Kim, J. H., Wise, J. H., and Abel, T. 2009, ApJ 694:123L\\
Klypin, A., Zhao, H. S., and Somerville, R. S. 2002, ApJ 573:597\\
Krick, J. E. and Bernstein, R. A. 2007, AJ 134:466\\
Kuntz, K. D. and Snowden, S. L. 2000, ApJ 543:195\\
Law, D. R. et al. 2007, ApJ 656:1\\
Li, Y. S. and White, S. D. M. 2008, MNRAS 384:1459\\
Li, Z. et al. 2006, MNRAS 371:147L\\
Li, Z., Wang, Q. D., and Hameed, S. 2007, MNRAS 376:960L\\
Loeb, A. et al. 2005, ApJ 633:894L\\
Manchester, R. N. et al. 2006, ApJ 649:235\\
Mannucci, F. et al. 2005, A\&A 433:807\\
Marconi, A. and Hunt, L. K. 2003, ApJ 589:21L\\
Martin, C. 2005, ApJ 621:227\\
Mateus, A. et al. 2006, MNRAS 370:721\\
Mathur, S., Weinberg, D. H, and Chen, X. 2003, ApJ 582:82\\
Maughan, B. J. et al. 2008, ApJS 174:117\\
McGaugh, S. S. et al. 2000, ApJ 533:L99\\
McGaugh, S. S. 2005, ApJ 632:859\\
Mo, H. J. and Mao, S. 2002, MNRAS 333:768\\
Murray, N., Quataert, E., and Thompson, T. A. 2005, ApJ 618:569\\
Navarro, J. F., Frenk, C. S., and White, S. D. M. 1997, ApJ 490:493\\
Navarro, J. F. 1998, preprint (astro-ph/9807084)\\
Navarro, J. F. and Steinmetz, M. 2000, ApJ 538:477\\
Nicastro, F. et al. 2002, ApJ 573:157\\
Nicastro, F. et al. 2005a, Nature 433:495\\ 
Nicastro, F. et al. 2005b, ApJ 629:700\\
Oppenheimer, B. D. and Dav\'{e}, R. 2008, MNRAS 387:577\\
Penton, S. V., Stocke, J. T., and Shull, J. M. 2004, ApJS 152:29\\
Pettini, M. et al. 2001, ApJ 554:981\\
Rasmussen, A., Kahn, S. M., and Paerels, F. 2003, ASSL Conf. Proc. 281:109\\
Rasmussen, J. et al. 2009, ApJ 697:79\\
Reid, M. et al. 2009, ApJ 700:137\\
Sakamoto, T., Chiba, M, and Beers, T.C. 2003, A\&A 397:899\\
Sembach, K. R. et al. 2001, ApJ 561:573\\
Sembach, K. R. et al. 2004, ApJS 155:351\\
Shattow, G. and Loeb, A. 2009, MNRAS 392:L21\\
Smith, R. K. et al. 2001, ApJ 556:L91\\
Snowden, S. et al. 1998, ApJ 493:715\\
Stanimirovi\'{c}, S. et al. 2002, ApJ 576:773\\
Stark, D. V., McGaugh, S. S., and Swaters, R. A. 2009, AJ 138:392\\
Strateva, I. et al. 2001, AJ 122:1861\\
Strickland, D. K. et al. 2004, ApJS 151:193\\
Sutherland, R. R. and Dopita, M. A. 1993, ApJS 88:253\\
Tang, S. et al. 2009, MNRAS 392:77\\
Tortora, C. et al. 2009, MNRAS 396:1132\\
T\"{u}llmann, R. et al. 2006, A\&A 457:779\\
van der Kruit, P. C. 1984, A\&A 140:470\\
Wang, Q. D. et al. 2005, ApJ 635:386\\
Wu, K. K. S, Fabian, A. C., and Nulsen, P. E. J. 2001, MNRAS 324:95\\
Xue, X. X. et al. 2008, ApJ 684:1143\\

\begin{figure}
\epsscale{1.0}
\plotone{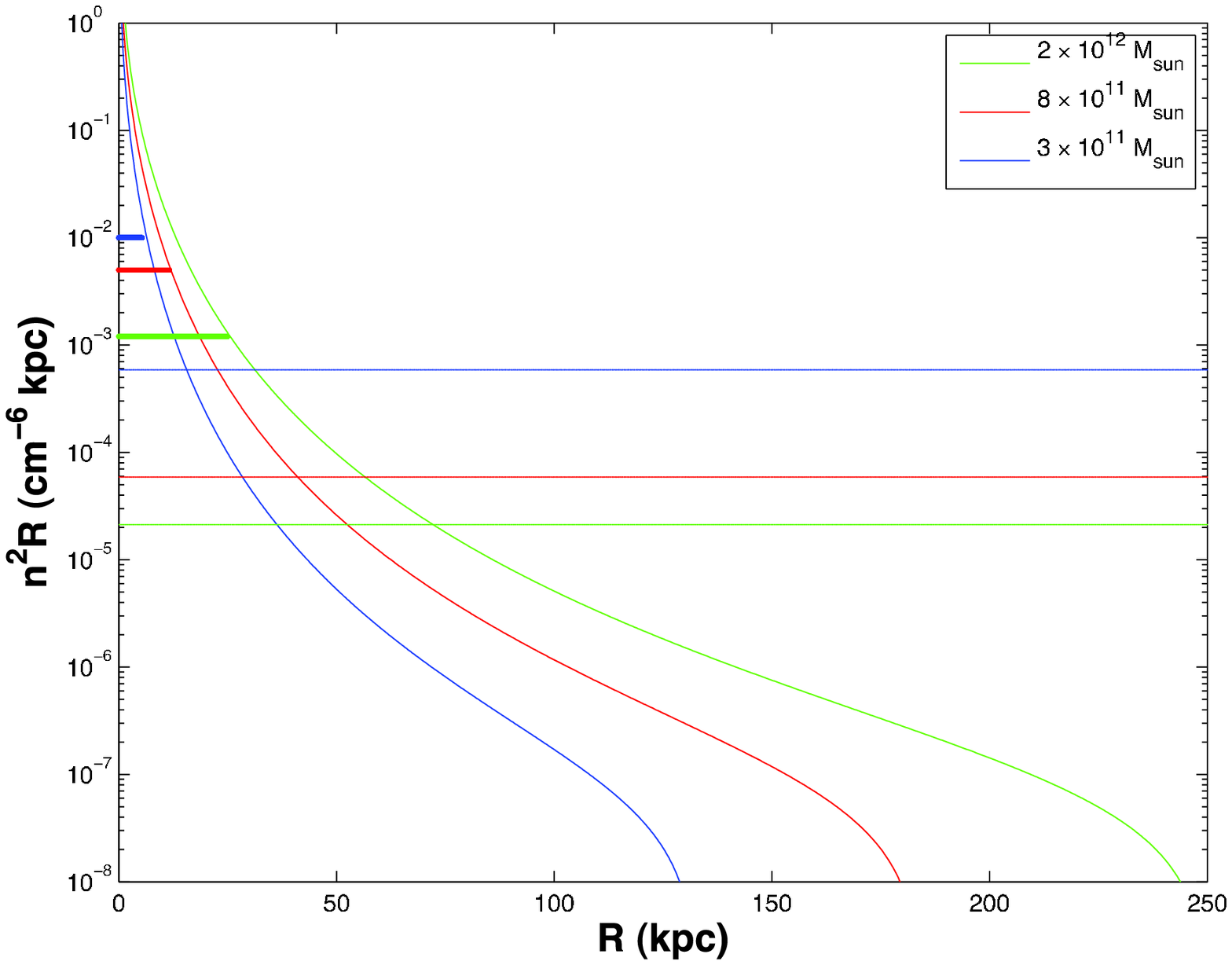}
\caption{\small Integrated X-Ray emission measures for a halo of hot gas obeying an NFW profile in hydrostatic equilibrium. The three colors denote the total mass of the host galaxy and its dark matter halo; each line represents the EM for an NFW profile of concentration parameter C=12. The three thick horizontal bars represent the approximate radial extent of the thick disk of a galaxy of the given mass; within these radii, the bright emission should be obscured and unobservable. The three thin horizontal lines are estimates of the minimum emission measures detectable with {\it Chandra} ACIS-S for each halo. These haloes are all observable out to many times the radius of the thick disk, and should be detected unless the halo mass is reduced to 11-24\% of the mass of missing baryons. The slope of the surface brightness between the thick disk and the detection limit follows a $\beta$-model with $\beta\approx 0.87$ in each case. }
\end{figure}

\begin{figure}
\epsscale{1.0}
\plotone{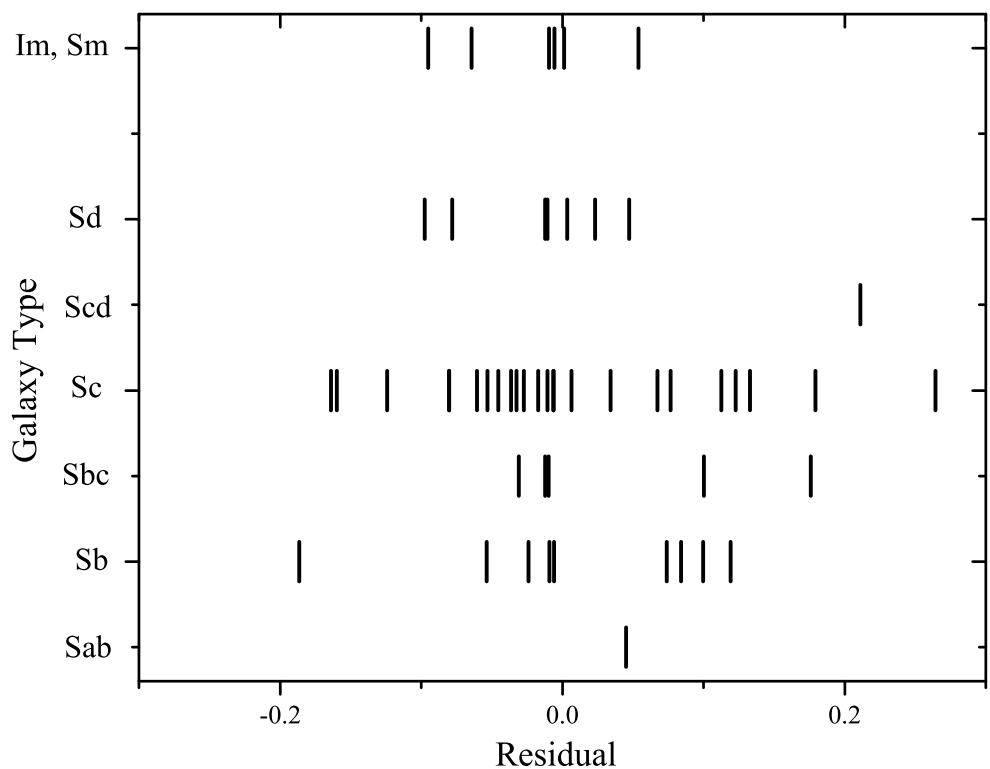}
\caption{\small  Comparison of the scatter of galaxies around the baryonic Tully-Fisher relation (BTF), based on the sample from McGaugh (2005). We use the galaxy type as a proxy for the size of the  supermassive black hole (SMBH). For the 60 galaxies with good galaxy types, there is no trend of the residual around the BTF with galaxy type. As discussed in section 6, we conclude that, at least for galaxies of type Sb and later, AGN activity associated with SMBHs does not play a dominant role in the removal of baryons from those galaxies.}
\end{figure}

\begin{figure}
\epsscale{1.0}
\plotone{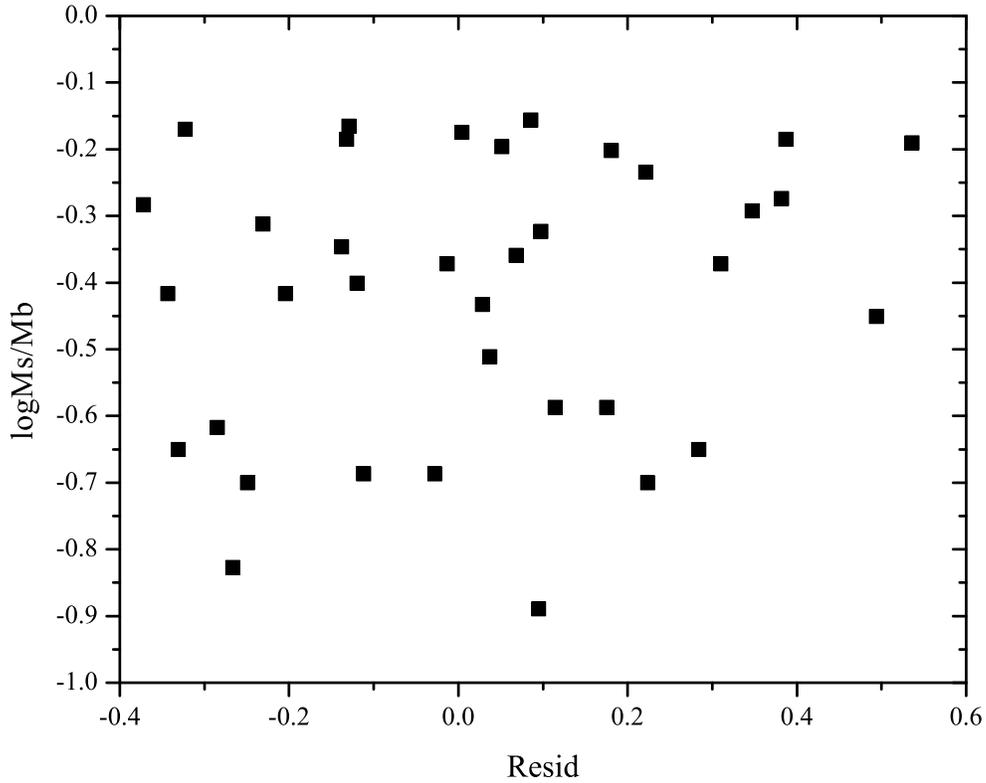}
\caption{\small Comparison of galaxy residuals about the baryonic Tully-Fisher relation as a function of their stellar mass fraction for the gas-dominated sample of Stark et al. (2009). For several of these galaxies, the stars comprise less than 25\% of the baryons, yet they have the same baryon fraction as the more star-rich galaxies. As discussed in section 6, we conclude that supernovae from the stars in these galaxies are not responsible for ejecting most of the missing baryons from the galaxies. }
\end{figure}

\begin{figure}
\epsscale{0.5}
\plottwo{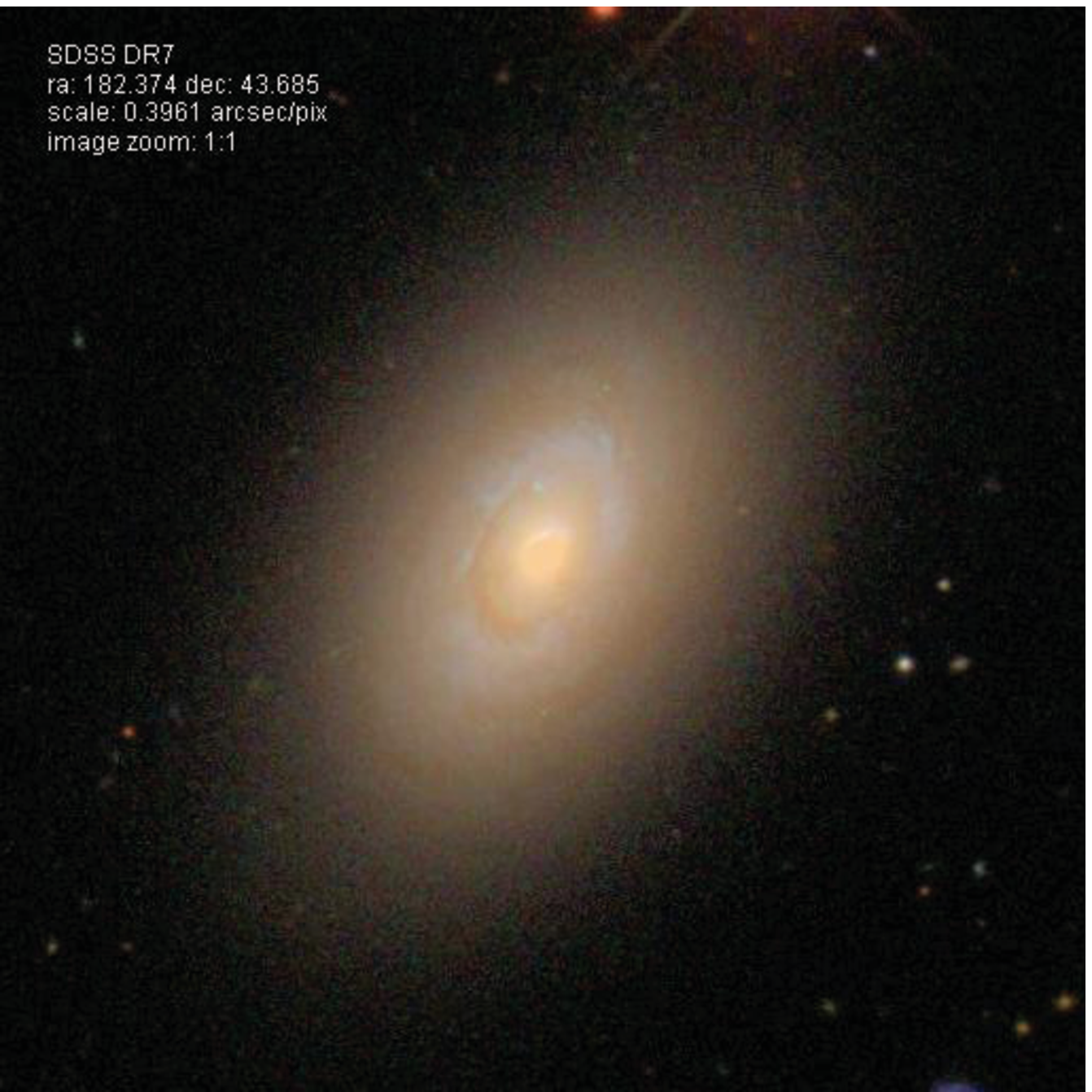}{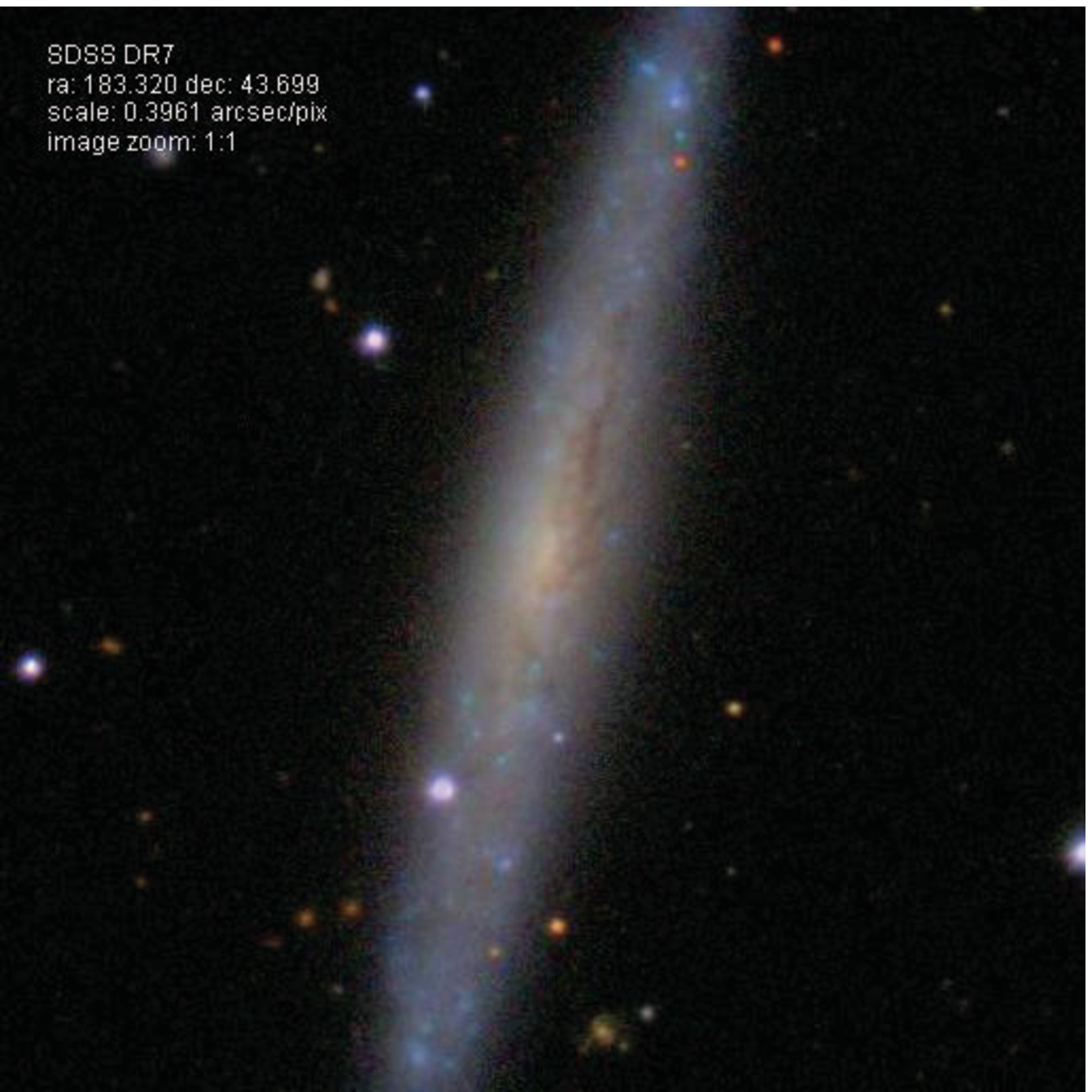}
\caption{\small SDSS images of NGC 4138 (left) and NGC 4183 (right). NGC 4138 is at a distance of $15.4\pm2.3$ Mpc, NGC 4183 is at a distance of $17.3 \pm 0.9$ Mpc. Both images have a scale of 0.3961 ''/pixel. For more information on and discussion of these galaxies, please see section 6.}
\end{figure}

\begin{figure}
\epsscale{1.0}
\plotone{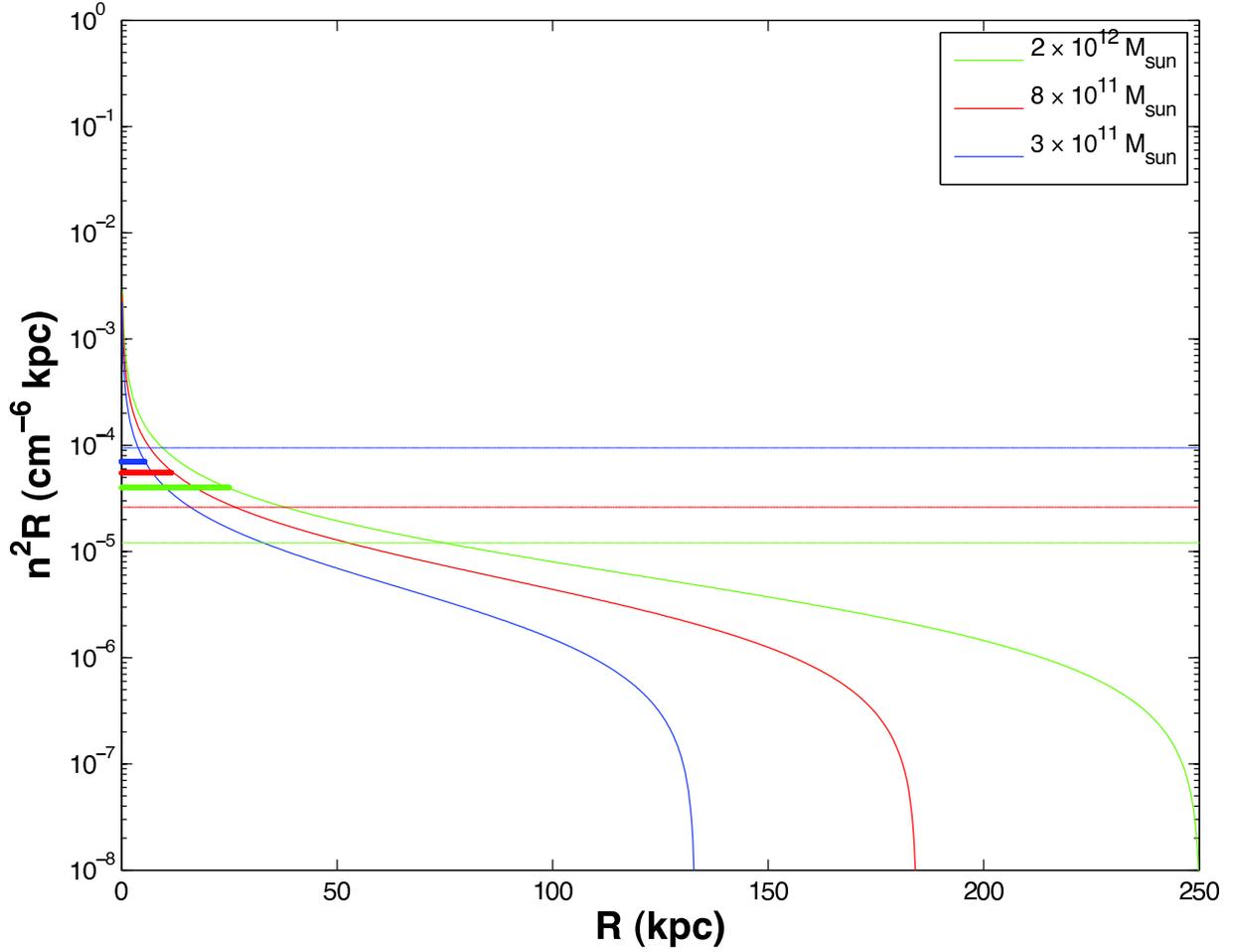}
\caption{\small As in Figure 1, the integrated X-Ray emission measures for a halo of hot gas, but in this plot the gas obeys a power-law profile: $n(r) \propto r^{-0.9}$. These hot haloes have a lower emission measure, but a higher emissivity per gram due to their increased temperature. The largest should be detected unless the halo mass is reduced to 59\% of the mass of missing baryons, the smaller haloes have weaker constraints. The slope of the surface brightness between the thick disk and the detection limit follows a $\beta$-model with $\beta\approx 0.36$ in each case. }
\end{figure}

\end{document}